\documentclass{elsart}

\begin{document}
\begin{frontmatter}
\centerline{Discrete Applied Mathematics }
  \title{Linear Visualization of a Directed Finite Graph with Labels on Edges}
\author{A.N. Trahtman, T. Bauer \and N. Cohen}

\date{}
%\ead{trakht@macs.biu.ac.il}
 \address{Bar-Ilan University, Dep. of Math., 52900, Ramat Gan, Israel}

\begin{abstract}
 Visualization has become an integral part in a wide array of application areas,
and papers are solicited both for original visualization research and for the
application of visualization towards the understanding of domain-specific data.

 Our main objective is the visual representation of a directed finite graph
with labels attached to edges based on the structure properties of the graph.
In particular, the transition graph of any deterministic finite automaton is also
 accepted.
A representation problem of a visual image of a digraph has
appeared in the study of the road coloring conjecture.
A polynomial-time algorithm for the road coloring
has been based on recent positive solution of this old famous
problem.

Our visualization program is used for demonstration of the algorithm
as well as for visualization of the transition graph of any deterministic
finite automaton ($DFA$). This help tool is linear in the size of the automaton.
\end{abstract}

\begin{keyword}
algorithm, visualization, road coloring, synchronization,graph, DFA
\end{keyword}

\end{frontmatter}

\section*{Introduction}
The visualization has become essential in many application areas.
The finite graphs and automata undoubtedly belong to such areas.
The visual perception of the structure properties of automata is
an important goal.

 Crucial role in the visualization plays for us the correspondence of the layout
 to the human intuition, the perception of the structure properties of the
graph and the rapidity of the appearance of the image. The
automatically drawn graphical image must resemble the last one of
a human being and present the structure of the graph. We use and
develop for this aim some known approaches \cite{ST}, \cite{WEK}.

Our algorithm for the visualization is linear in the size of the
automaton. This not complicated at first sight algorithm
successfully solves a whole series of tasks of the disposal of the
objects. The pictorial diagram demonstrates the graph structure
highlighting strongly connected components, paths and cycles.
So this kind of visualization can be considered as a structure
visualization.

A problem of a visual image of a directed finite graph has appeared
in the study of the {\em road coloring conjecture}.
The conjecture  \cite{AW},  \cite{AGW}, \cite{Ru}
 was stated about forty yeas ago for a complete strongly connected directed
finite graph with constant outdegree of all its vertices where the greatest
 common divisor (gcd) of lengths of all its cycles is one.
The edges of the graph being unlabelled, the task is to find a
labelling that turns the graph into a deterministic finite
automaton possessing a synchronizing word. According to the
conjecture, such a graph has a synchronizing coloring.

The problem belonged to the most fascinating problems
in the theory of finite automata  \cite{MS},  \cite{CKK}  and was mentioned
 in the popular Internet Encyclopedia "Wikipedia" on the list of the
 interesting unsolved problems in mathematics.
The positive solution of the road coloring problem  \cite{Tc}, \cite{Ti}
 is a basis of a polynomial-time implemented algorithm of $O(n^3)$ complexity
in the worst case.

The realization of the considered algorithm is demonstrated by a high-speed
visualization program. The visualization of the transition graph of the automaton
is, in particular, a help tool of the study of automata. Thus the linearity of the algorithm
is comfortably and important. Both the road coloring algorithm and the visualization
 algorithm are implemented in the package TESTAS (www.cs.biu.ac.il/$\sim$trakht/syn.html).

The visibility of inner structure of a digraph without doubt is a matter of
interest not only for the road coloring, the range of the application may be
significantly wider and includes all directed graphs with labels on edges.

\section*{Graphs, automata and visualization}
A {\it directed graph} or {\it digraph} is a pair $G = (V,E)$ of a set $V$,
 whose elements are called {\it vertices} and of a set $E$ of ordered pairs of vertices,
 called {\it edges}. Graphs with labels (colors) attached to edges are designated as labelled.

A {\it path} in a digraph $G$ is a sequence of edges $e_1,...,e_k$ such that
the end vertex of $e_i$ is the start vertex of $e_{i+1}$ for $ i=1,2,...,k-1$.
The path is called a {\it cycle} if $e_1=e_k$.

As usual, we regard a directed graph with colors assigned to its edges as a finite
automaton, whose input alphabet consists of these colors. The graph is called
{\it transition graph} of the automaton.

An automaton is {\it deterministic} if no state has two outgoing edges
of the same color. In {\it complete} automaton each state has outgoing
edges of any color.

If there exists a path in an automaton from the state $\bf p$ to
the state $\bf q$ and the edges of the path are consecutively
labelled by $\sigma_1, ..., \sigma_k \in \Sigma$, then for
$s=\sigma_1...\sigma_k \in \Sigma^+$ let us write  ${\bf q}={\bf
p}s$.

 Let $|P|$ denote the size of the subset $P$ of states from an automaton
  (of vertices from a graph).

Let $Ps$ be the set of states ${\bf p}s$ for ${\bf p} \in P$
 $s \in \Sigma^+$. For the transition graph
$\Gamma$ of an automaton let $\Gamma s$
denote the map of the set of states of the automaton.

 A word $s \in \Sigma^+ $ is called a {\it synchronizing}
word of the automaton with transition graph $\Gamma$
if $|\Gamma s|=1$.

 A coloring of a directed finite graph is {\it synchronizing} if the
coloring turns the graph into a deterministic finite automaton
possessing a synchronizing word.

We call the set of all outgoing edges of a vertex a {\it bunch} if
 all these edges are incoming edges of only one vertex.

Imagine a map with roads which are colored in such a way that a
fixed sequence of colors, called a synchronizing sequence, leads
to a fixed place whatever is the starting point. Finding such a
coloring is called {\em road coloring problem}. The roads of the map are
considered as edges of a directed graph. The visual presentation
of a road coloring algorithm is essentially based on the paths of
the graph. The paths must be visible as well as cycles, bunches
and other structure components of the graph. In particular, the
notion of the bunch according to the following lemma plays some role in
the road coloring algorithm.
\begin{lem}  $\label {f7}$ \cite{Ti}
 If some vertex of graph $\Gamma$ has two incoming bunches then
there exists a stable pair by any coloring.
 \end{lem}
The role of maximal length path in a subtree is also
important. These maximal subtrees are in the center of some proofs,
for instance:
\begin{lem}  $\label {f9}$ \cite{Ti}
Let any vertex of the graph $\Gamma$ have no two incoming bunches.
Then $\Gamma$ has a subgraph of some color with single maximal subtree.
 \end{lem}
\begin{lem}  $\label {f11}$ \cite{Tc}
Let any edge of given color belong to a cycle of edges of this
color. Then $\Gamma$ has a coloring of some color with single
maximal subtree.
 \end{lem}
 The road coloring problem was stated for a strongly connected directed finite
graph \cite{AW} because of the key role of the strongly connected
components ($SCC$) of the graph for the problem. Therefore the
visual image of $SCC$ is highly important for study of road
coloring.

\section{An approach to the visualization}

 A crucial role in the visualization plays in our opinion the correspondence of the layout
to the human intuition, the perception of the structure properties of the
graph and the rapidity of  the appearance of the image. The automatically drawn
graphical image must resemble the last one of a human being.
The considered visualization is a help tool for any program dealing with
transition graph of $DFA$ and in particular for the road coloring algorithm.

Our main objective is a visual representation of a directed graph with labels
on its edges and, in particular, of the transition
graph of a deterministic finite automaton based on the structure
properties of the graph. The pictorial diagram demonstrates the
graph structure.
 Among the important visual objects of a digraph one can mention paths,
cycles, strongly connected components, cliques, bunches etc. These
properties reflect the inner structure of the digraph.
The strongly connected components ($SCC$)  are here of special significance.

We choose here a cyclic layout \cite{ST}, \cite{WEK}. According to
this approach the vertices are placed at the periphery of a
circle. Our modification of the approach considered two levels of
circles, the first level consists of strongly connected
components, the second level corresponds to the
 whole graph with $SCC$ at the periphery of the circle. The visual placement
is based on the structure of the graph considered as a union of the set
of strongly connected components.

 Clearly, the curve edges (used, for instance, in the package
GraphViz \cite{EGK}, \cite{Si}) hinder to recognize the cycles and
paths. Therefore, we use only direct and, hopefully, short edges.
We have changed some priorities of the layout and, in particular,
eliminate the goal of reducing the number of intersections of the
edges as it was an important aim in some algorithms \cite{STT},
\cite{Si}. The intersections of the edges are even not considered
in our algorithm. This approach gives us an opportunity to
simplify essentially the procedure and to reduce its complexity.
Our main intent is only not to stir by the intersections of the
edges to conceive the structure of the graph. The intersections
are placed in our algorithm far from the vertices due to the
cyclic layout \cite{WEK}, \cite{ST} we use. The area of vertices
differs of the area of the majority of intersections.

\section{Visualization algorithm}

The strongly connected components (SCC) are of special
significance in the algorithm. Thus our first step is the eduction
and selection of the $SCC$. The quick linear algorithm for finding
$SCC$ \cite {AHU} is implemented in the program. The radius of the
big cycle is a function of diameters of $SCC$ cycles. All $SCC$
are placed on the periphery of a big circle and are ordered
according to the size (the number of vertices).

 So strongly connected components can be easily recognized by the observer.
The pictorial diagram demonstrates the structure of the graph and
 the visualization can be considered as a kind of a structure visualization.

According to the cyclic approach, the vertices arranged in a circle of
 $SCC$ in the graph layout.

The periphery of a circle of $SCC$ is the most desirable area for
placing the edges because the edges in this case are relatively
short. We choose the order of the vertices of the $SCC$ on the
circle according to this purpose. The length of some edges can be
reduced in a such way. It also helps to recognize paths and cycles
on the screen.

From the other hand, the edges between distinct
$SCC$ are relatively longer than the inner edges of strongly
connected components.

  The problem of the placing of the labels near corresponding edges
is sometimes very complicated and frequently the connection
between the edge and its label is not clear. Our solution is to use colors
on the edges instead of labels and exclude the placing of labels.

The set of loops of arbitrary vertex is placed around the vertex with
some shift that depends on the size of the set. The problem of parallel
edges is solved analogously, the origins of the edges must belong to the vertex.

 The linearity of the algorithm ensures the momentary appearance of
the layout. It is favorably also for educational purposes
 because the road coloring conjecture can be stated in simple terms and initial
explorations can be done immediately.
"It can be understood by any student with a little experience in the graph theory.
The Road Coloring Conjecture makes a nice supplement to any discrete mathematics
 course" \cite{Ra}.

The complexity of the algorithm shows the following

 \begin{lem}  $\label {p4}$
The time and space complexity of the visualization algorithm described above
is linear in the sum of states and edges of the transition graph of the automaton.
 \end{lem}
Proof. The algorithm for finding strongly connected components ($SCC$)
 of the graph is linear in the sum of states and edges \cite {AHU}.
 One can find in a linear time the order of vertices on a SCC circle
for to put more edges on the periphery of the circle. The step for
finding the places for all SCC in the big circle of the graph
together with its radius is linear in the number of $SCC$. The
placing of direct edge depends on its borders and does not change
the complexity.

The program uses only linear arrays.

  The algorithm is described in the following pseudocode

\subsection*{{\sc FindLayout}(deterministic automaton $A$)}

1\quad  Find list of $SCC$  $A_i$ of $A$

2\quad  Reorder $SCC$ $A_i$ according to the size

3\quad Divide the circumference of big cycle according to the sizes

4\quad Define the place of the  center of $A_i$ on the big circle

5 for each  $SCC$ $A_i$

6 \quad  do

7 \quad{    }   Find relatively great cycle of $A_i$

8 \quad{    }   Reorder states of $A_i$ for to place the cycle consecutively on its circle

9 \quad{    }   Place the states on the circle of $A_i$

10\quad     end do

11 for any state

12\quad{}     do

12  \quad{}         for any outgoing edge

13  \quad{}\quad{    }           check end of the edge

14  \quad{}\quad{    }            define the length of the edge

15  \quad{}     form subsets of parallel edges

16  \quad{}    if there are parallel edges

16  \quad{} \quad{} define shifts on the stripe of parallel edges

17   \quad{} place outgoing edges of the state

18  \quad{}     form set of loops

19  \quad{}    if there is more than one loop

20  \quad{  } \quad{  } define shifts for the placing of loops

21 \quad{} place loops of the state

22\quad{}      end do

23  return image

\section{Input of data and examples}
The transition graph of any deterministic finite automaton is
accepted by the visualization algorithm. The transitions graphs of
non-complete automata also can be reproduced. More generally, any
Cayley table is an input for the algorithm.

The input file is an ordinary txt file for all algorithms used in
the package TESTAS. We open source file with transition graph of
the automaton in the standard way and then check different
properties from menu bar. The graph is shown on the display by
help of a rectangular table. More precisely,
 transition graph of an automaton as well as an arbitrary directed graph
with distinct labels on outgoing edges of every vertex is presented by the matrix
(Cayley graph):

                       \centerline{vertices X labels}

   First two numbers in input file are the size of alphabet of labels
  and the number of vertices.  The integers from 0 to n-1 denote the
vertices.
  i-th row is a list of successors of i-th vertex according to the label in the
  column (number of the vertex from the end of  edge with label from the j-th column
  and beginning in i-th state is placed in the (i,j) cell).

  The User defines the data. He can define the number of nodes, size of
 the alphabet of edge labels and the values in the matrix.
 For example, the input 2 6 1 0 2 1 0 3 5 2 3 2 4 5
 presents the Cayley graph with 2 labels and 6 vertices and the next input
  2 5 1 0 2 1 ; 3 5 ; 3 ; presents the  Cayley graph with 2 labels and 5 vertices.
The values are divided by a gap.  The semicolon corresponds to
empty cell of the table.

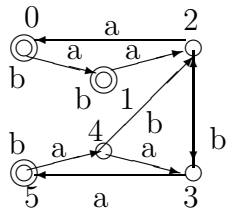
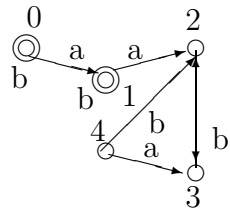
\begin{center}
\begin{tabular}{|c|c|c|}
 \hline
    & letter $ a $ & letter $ b $ \\ \hline
  $ vertex $ 0 & 1 & 0 \\ \hline
  $ vertex $ 1 & 2 &  1 \\ \hline
  $ vertex $ 2 & 0 &  3 \\ \hline
  $ vertex $ 3 & 5 &  2 \\ \hline
  $ vertex $ 4 & 3 & 2 \\ \hline
  $ vertex $ 5 & 4 &  5 \\
  \hline
\end{tabular}  $ and $
\begin{tabular}{|c|c|c|}
 \hline
    & letter $ a $ & letter $ b $ \\ \hline
  $ vertex $ 0 & 1 & 0 \\ \hline
  $ vertex $ 1 & 2 &  1 \\ \hline
  $ vertex $ 2 &  &  3 \\ \hline
  $ vertex $ 3 & 5 &  \\ \hline
  $ vertex $ 4 & 3 &  \\
  \hline
\end{tabular}
\end{center}

\begin{picture}(50,78)
\end{picture}
\begin{picture}(130,78)
\multiput(6,60)(64,0){2}{\circle{6}}
\multiput(6,13)(64,0){2}{\circle{6}}

\put(6,68){0}
 \put(66,67){2}
 \put(42,37){1}
 \put(6,0){5}
 \put(66,0){3}
 \put(30,25){4}

 \multiput(22,56)(22,0){2}{a}
\multiput(16,19)(34,0){2}{a}
 \put(36,21){\circle{6}}
\put(36,48){\circle{6}}
 \put(7,14){\vector(4,1){28}}
\put(7,57){\vector(4,-1){26}}

\put(39,52){\vector(4,1){27}}
 \put(37,20){\vector(4,-1){28}}
\put(67,63){\vector(-1,0){57}}
 \put(36,65){a}
\put(67,12){\vector(-1,0){57}}
 \put(32,0){a}

\put(70,15){\vector(0,1){42}}
 \put(70,59){\vector(0,-1){42}}
\put(34,21){\vector(1,1){36}}
 \put(52,28){b}

  \put(76,22){b}

\put(25,37){b} \put(36,48){\circle{10}} \put(0,20){b}
\put(0,45){b}

\put(6,60){\circle{10}}
 \put(6,13){\circle{10}}
 \end{picture}
\begin{picture}(45,78)
\end{picture}
\begin{picture}(130,78)
\multiput(6,60)(64,0){2}{\circle{6}}
 \put(70,13){\circle{6}}
 \multiput(22,56)(22,0){2}{a}
\put(50,19){a}
 \put(36,21){\circle{6}}
\put(36,48){\circle{6}}

\put(6,68){0}
 \put(66,67){2}
 \put(42,38){1}
 \put(66,0){3}
 \put(30,25){4}

\put(7,57){\vector(4,-1){26}}

\put(39,52){\vector(4,1){27}}
 \put(37,20){\vector(4,-1){28}}
%\put(67,63){\vector(-1,0){57}}
 %\put(36,65){a}

\put(70,15){\vector(0,1){42}}
\put(70,59){\vector(0,-1){42}}
\put(34,21){\vector(1,1){36}}
 \put(52,28){b}

  \put(76,22){b}

\put(25,37){b}
\put(36,48){\circle{10}}

\put(0,45){b}

\put(6,60){\circle{10}}

 \end{picture}

The road coloring algorithm changes some colors of edges of the
transition graph. An example of the repainting of
non-synchronizing graph follows (see also the appendix).

A non-synchronizing source graph. $\qquad \quad $ The repainted
synchronizing graph.

%\begin{center}
\begin{tabular}{|c|c|c|}
 \hline
    & letter $ a $ & letter $ b $ \\ \hline
  $ vertex $ 0 & 6 & 2 \\ \hline
  $ vertex $ 1 & 5 &  3 \\ \hline
  $ vertex $ 2 & 6 &  6 \\ \hline
  $ vertex $ 3 & 1 &  1 \\ \hline
  $ vertex $ 4 & 1 & 1 \\ \hline
  $ vertex $ 5 & 7 & 7 \\ \hline
  $ vertex $ 6 & 8 & 8 \\ \hline
  $ vertex $ 7 & 0 & 0 \\ \hline
  $ vertex $ 8 & 7 & 5 \\ \hline
  $ vertex $ 9 & 4 & 8 \\
  \hline
\end{tabular}  $ repainting ->$
\begin{tabular}{|c|c|c|}
 \hline
    & letter $ a $ & letter $ b $ \\ \hline
  $ vertex $ 0 & 2 & 6 \\ \hline
  $ vertex $ 1 & 5 &  3 \\ \hline
  $ vertex $ 2 & 6 &  6 \\ \hline
  $ vertex $ 3 & 1 &  1 \\ \hline
  $ vertex $ 4 & 1 & 1 \\ \hline
  $ vertex $ 5 & 7 & 7 \\ \hline
  $ vertex $ 6 & 8 & 8 \\ \hline
  $ vertex $ 7 & 0 & 0 \\ \hline
  $ vertex $ 8 & 7 & 5 \\ \hline
  $ vertex $ 9 & 4 & 8 \\

  \hline
\end{tabular}
%\end{center}

 The minimal length
synchronizing word (found by the package TESTAS) is now
$aaaabbaaabba$. Two edges have changed the color.

There are known some automata with the maximal length of the minimal
synchronizing word for $n$-state automaton  found by \v{C}erny
\cite{Ce}: 1)the infinite sequence of \v{C}erny, 2)eight single
automata (see \cite{Kr}, \cite{Ro},  \cite{Ta}, \cite{CPR}). These
graphs are extreme because it is a full list of all known 
 automata with a minimal length of its synchronizing word
equal to the maximal value $(n-1)^2$.  The visualization program
demonstrates this remarkable set of eight automata together  with
one automaton of size eight from the \v{C}erny sequence (see
appendix). The input of the program was a table of size 40X3. This
set of graphs demonstrates two levels of cyclic layout. The big
cycle is filled by strongly connected above-mentioned graphs, the
vertices of any such graph also located on a corresponding little
cycle.

       %\centerline{The appendix}
\end{document}